\documentclass[reprint,showpacs,amsmath,amssymb,aps,10pt]{revtex4-1}
\usepackage{graphicx}
\usepackage{bm}
\usepackage[breaklinks=true,colorlinks=true,linkcolor=blue,urlcolor=blue,citecolor=blue]{hyperref}
\usepackage{graphics}
\usepackage{dcolumn}
\usepackage{amsmath,amssymb}
\usepackage{mathdots}
\usepackage{natbib}
\usepackage{soul,color}
\usepackage{epstopdf}

\begin{document}
\title{$^4$He sample probe for combined microwave and dc transport measurements}

\author{Oleksandr V Dobrovolskiy$^{1,2}$, J\"org Franke$^1$, and Michael Huth$^1$}
\address{$^1$ Physikalisches Institut, Goethe University, 60438 Frankfurt am Main, Germany}
\address{$^2$ Physics Department, V. Karazin National University, 61077 Kharkiv, Ukraine}
\begin{abstract}
Combined microwave and dc electrical transport measurements at low temperatures represent a valuable experimental method in many research areas. In particular, when samples are conventional superconductors, a typical experiment requires a combination of helium temperatures, a wide range of magnetic fields, and the utilization of coaxial lines along with the usual dc wiring. We report on the general design features and the microwave performance of a custom-made low-temperature sample probe, with a measurement bandwidth tested from dc to 20~GHz. Equipped with 6 coaxial cables, a heater, Hall and temperature sensors, the probe fits into a $\oslash32$~mm shaft. We present our setup, analyze its microwave performance, and describe two representative experiments enabled by this system. The proposed setup will be essential for a systematic study of the dc and ac response of the vortex dynamics in nanopatterned superconductors subject to combined dc and microwave stimuli. Besides, it will be valuable for the investigation of a broad class of nonlinear stochastic systems where a combination of dc and high-frequency ac driving in a wide temperature range is necessary.
\end{abstract}

\date{February 01, 2015}

\maketitle

\section{Introduction}\label{sIntro}
Combined microwave and dc electrical transport measurements represent a valuable experimental method in many research domains. These include investigations of Josephson junctions~\cite{Sha63prl,Van85prb,Knu12pre}, quantum computing~\cite{Ong12rsi,Cla08nat}, magnetization reversal~\cite{Lar14apl}, phase slips in superconducting bridges~\cite{Siv03prl,Dmi07sst}, and the light-matter interaction on a chip~\cite{Nie10nph,Wal04nat}. Further studies deal with electrical transport and pinning properties of crystals~\cite{Zho13prb,Bae08prb}, mesoscopic devices and nanowires~\cite{Hon12prl,Web13prb}, and the dynamics of charge density~\cite{Zyb13prb} and spin density~\cite{Bar93prl} waves. In these experiments, microwaves are needed for probing the relevant energy/time/length scale of the system, while the quantum effects become relevant at low temperatures.

An appealing subject for combined dc and microwave measurements is the dynamics of Abrikosov vortices in type II superconductors~\cite{Git66prl,Fio71prl,Mar76prl,Har95prl,Kok02prl,Pom13apl,Jin10prb,Son09apl,Oku07prb,Von11prl,Wor12prb,Sol14prb,Awa11prb}. In particular, when fluxons coherently move in a periodic pinning landscape under superimposed dc and ac drives, synchronization effects become apparent through Shapiro steps~\cite{Sha63prl} in the current-voltage curve (CVC)~\cite{Fio71prl,Mar76prl,Bae08prb,Von11prl} and mode-locking peculiarities in the complex impedance response~\cite{Pom08prb} and the absorbed microwave power~\cite{Jin10prb,Wor12prb}. In general, a superconductor with a transition temperature $T_c$ of $10$~K is characterized by a zero-temperature gap frequency $f_B(0)\equiv2\Delta(0)/h$ of the order of $100$~GHz, above which the superconducting state is destroyed due to gap breakdown. Therefore, the measurement of the dc voltage and the absorbed microwave power at frequencies well below the gap frequency represents a powerful approach for studying the pinning mechanisms and the dissipation processes in type II superconductors. Accordingly, recent literature traces growing interest in the investigation of vortex matter in nanostructured superconductors~\cite{Mos10boo,Mos11boo}, with particular focus on studying the flux transport mechanisms at microwaves~\cite{Wor12prb,Sol14prb,Awa11prb}. Whereas a number of new effects in the two-dimensional, microwave-driven dynamics of vortices have been predicted recently~\cite{Shk08prb,Shk11prb,Shk14pcm}, there have been few experiments to examine these in high-$T_c$ superconductors~\cite{Wor12prb} and virtually no experiments have been performed on low-$T_c$ superconductor thin films so far. This has motivated us to design the high-frequency sample probe presented here.

Typical issues one faces when designing a sample probe for an experiment in the radiofrequency and the lower microwave ranges are the following: First, along with the thermo\-metry-related and conventional dc wiring, the probe should be equipped with coaxial cables for handling signals with a frequency up to a few tens of GHz. Second, in the case of low-$T_c$ superconductors, the probe should be suitable for measurements from room down to helium temperatures. Third, care must be taken for matching the impedance of the sample under study with that of the transmission line. Fourth, since measurements of the mixed state assume a magnetic field, the sample space is usually limited by the solenoid's inner hole diameter or the coil diameter. At the same time, the thinner the coaxial cables employed, the higher is the attenuation for a given frequency. A last but not least issue is that the heat leakage due to the transmission line should be minimized.

The literature traces a considerable number of reports on the design and performance of various microwave sample probes and systems. We refer to~\cite{Ste12rsi,Boo94rsi,Ong12rsi,Ave14arx,Liu14rsi,Kit08rsi,Jin10prb,Wor12prb,Awa11prb,Sol14prb} to name just a very minor portion of them. In particular, these systems include apparatuses designed to work at $^3$He temperatures~\cite{Ong12rsi,Ste12rsi,Ave14arx,Liu14rsi} and a series of multi-port sample housings~\cite{Ong12rsi,Jin10prb,Wor12prb,Ave14arx}. A notable portion of works relates to the Corbino disk geometry~\cite{Ste12rsi,Boo94rsi,Pom10pcs,Liu14rsi,Kit08rsi}, other are concerned with coplanar waveguides~\cite{Ong12rsi,Ave14arx,Awa11prb}, whereas reports on the combined dc and broadband microwave measurements of the vortex dynamics in the microstrip geometry remain rare.~\cite{Wor12prb}.

Here, we present a custom-made insertable six-port $^4$He sample probe for combined broadband microwave and dc electrical transport measurements. Our setup successfully combines several well-established features present in other setups, refer to Table~\ref{Table}, and goes beyond their limits in selected characteristics. Namely, as our sample probe has been specifically developed for exploring the dynamics of Abrikosov vortices in thin-film superconductors its features have been optimized to meet the respective needs as follows. (i) The setup covers the temperature range from room temperature to 1.8 K and the magnetic field range to 14 T. (ii) Electrically, the probe has been tested for superimposed dc and ac stimuli from $0.3$~MHz to $20$~GHz and has shown a reliable performance over a two-year operation. (iii) Mechanically, the designed construction fits into a $\oslash32$~mm cryostat shaft, that is quite compact given that system is suitable for six-channel measurements. The availability of many channels is fortunate for the development and investigation of sophisticated microwave devices. (iv) Finally, the presented setup is versatile due the detachable sample housing on the cold end of the sample probe allowing the proposed assembly to be used for a broad range of experiments.
\begin{table}
\footnotesize{
\begin{tabular}{l*{4}{c}r}
Ref.                    &   $f$, GHz         & temperature                   & geometry      &  No      \\ \hline
\cite{Jin10prb}      &  2                    & $^4$He -- RT                 & stripline       & 4              \\
\cite{Wor12prb}    &  8                    & N$_2$ -- RT                    & stripline       & 4             \\
\cite{Awa11prb}    &  8                    & $^4$He -- RT                 & stripline       & 2              \\
\cite{Ong12rsi}     &  12                   & 30\,mK $\lesssim$ 0.1\,K  & stripline        & 6              \\
\cite{Ave14arx}    &  32                    & 20\,mK -- ?                   & stripline       & 4              \\
\cite{Kit08rsi}      &  10                    & 10 K -- RT                    & Corbino disk  & 1              \\
\cite{Liu14rsi}     &  16                    & 300\,mK -- 6\,K             & Corbino disk   & 1              \\
\cite{Ste12rsi}     &  20                    & 450\,mK -- 2\,K             & Corbino disk  & 1              \\
\cite{Boo94rsi}     &  20                    & $^4$He -- RT               & Corbino disk  & 1              \\
this work             &  20                   & 1.8\, K -- RT                 & stripline$^\ast$& 6             \\
\end{tabular}
}
\caption{Selected features of our sample probe compared with some presented in the literature. The upper frequencies and the experimental geometries are drawn from the experimental data reported in the cited references. Here $f$ is the high-end (tested) working frequency of the setup, ``RT'' means room temperature, ``stripline'' stands for a planar transmission line in its various modifications (coplanar waveguide, hairpin line, microstrip line etc.), while ``$^4$He'' and ``N$_2$'' denote helium and nitrogen temperatures, respectively. $^\ast$Though in what follows we present the results of measurements in the microstrip geometry only, other geometries are also possible with our sample probe due to the detachable sample housing.}
\label{Table}
\end{table}

The paper is organized as follows. The mechanical assembly and the design features of the setup are addressed in Sec.~\ref{cMechanical}. The microwave performance of the sample probe is analyzed in Sec.~\ref{cMicrowave}. Two exemplary experiments are presented in Sec.~\ref{cNiobium}. Conclusions round up our presentation in Sec.~\ref{cConclusion}.

\section{Mechanical assembly}\label{cMechanical}
\begin{figure*}
    \centering
    \includegraphics[width=0.8\textwidth]{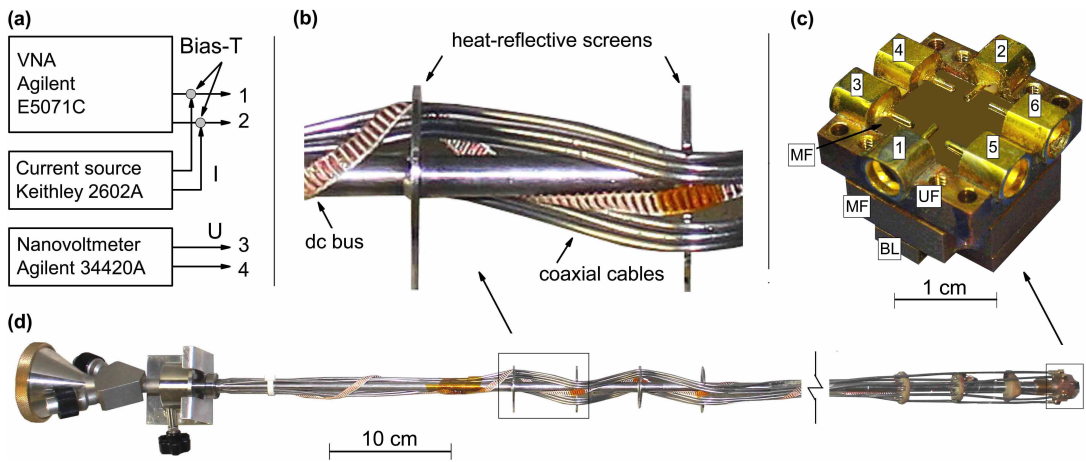}
    \caption{Sketch of the experimental setup. (a)~The instrumentation test set for sourcing and analyzing combined dc and microwave signals. (b)~Six coaxial cables and the dc bus passing by the heat-screening shields of  the sample probe. (c)~The detachable sample housing box, the top lid unmounted. (d)~Photoimage of the completely equipped top-loading $^4$He sample probe.
    \label{fProbe}}
\end{figure*}

The experimental setup is schematically shown in Fig.~\ref{fProbe}. It consists of the following major parts. (a) The instrumentation test set is placed right next to the top of the cryostat. In the sketch, standard temperature- and magnetic field-related instrumentation is not shown. (b) The custom-made insertable sample probe is capable of delivering superimposed dc and microwave stimuli. (c) The detachable sample-housing box is mounted on the cold end of the sample probe and determines the geometry of a particular experiment. We consider these parts in detail next.

(a) The microwave signal is generated by an Agilent E5071C vector network analyzer (VNA). The VNA is placed in an instrumentation rack as close as possible to the top of the cryostat. This is done for keeping the overall length of the microwave transmission line as short as possible and, thereby, for reducing the damping of the microwave signal. The VNA is connected to the sample probe by a pair of flexible coaxial cables with a length of 1.2~m. Two detachable bias tees~\cite{Pasternack} are mounted right at the ports of the VNA. They serve for superimposing the applied dc and microwave signals and for uncoupling those passed through the sample under study. The particular VNA is equipped with a $300$~KHz$-20$~GHz four-port test-and-measurement kit and is capable of supplying a maximal microwave power of $+10$~dBm. The frequency pass band in the ``ac arm" of the bias tees ranges from $100$~KHz to $18$~GHz.

(b) The sample probe is placed in a variable temperature insert (VTI) of an Oxford Instruments helium-flow cryostat. A set of heaters and sensors in the VTI nominally allows the temperature to be set to any value from 1.5 to 300~K, with a stability within $\Delta T = 1$~mK. The cryostat is equipped with a $14$~T superconducting solenoid. For the sample probe we used an un-wired $\oslash8$~mm stainless steel rod ordered from the cryostat manufacturer. The insertable section of the sample probe is 1.5~m in length, which allows the sample to be placed in the geometrical center of the solenoid. For the temperature and magnetic field sensing a bus of dc wires has been wrapped around the rod. Care has been taken for reducing the thermal leakage by radiation along the rod by mounting a set of heat-reflective screens. For handling the microwave signal, the probe is equipped with 6 semi-rigid coaxial cables~\cite{Coaxes} with a length of 1.7~m and an outer conductor diameter of 2.2~mm. The outer conductor is made from stainless steel 304, while the center conductor is made from a beryllium-copper alloy. The frequency range of the employed semi-rigid coaxial cables is rated up to 61~GHz. Nominal room-temperature insertion losses at frequencies of 1~GHz and 10~GHz are 1.46~dB/m and 4.79~dB/m, respectively. The coaxial cables are arranged essentially along the rod with the exception of the places where these have to pass by the heat-reflective shields, as is shown in Fig.~\ref{fProbe}(b). On the warm end of the sample probe all coaxial cables are soldered to the vacuum-sealed SMA connectors, while on its cold end they are soldered to the right-angle mini-SMP connectors~\cite{Rosenberger}. The sample probe thus equipped, see Fig.~\ref{fProbe}(d), fits into the $\oslash32$~mm VTI shaft.

(c) The detachable sample-housing copper box is screwed on the cold end of the sample probe. Its lateral dimensions are limited by the solenoid inner hole diameter. The housing consists of the main frame~(MF), onto which the sample is mounted, the upper frame~(UF), as well as the bottom~(BL) and top lids, see Fig.~\ref{fProbe}(c). Six mini-SMP connectors are precisely soldered onto UF which has a square inner space with a side of $10.5$~mm. Accordingly, the housing allows for accommodating samples with lateral dimensions of up to $10\times10$~mm$^2$. The free volume inside the housing is kept as small as possible to move any box resonances above $20$~GHz. Our experiment is primarily oriented on measurements on thin films in the microstrip geometry so that this will be in the focus of our subsequent presentation, while hair-pin, meander, stripline, and coplanar waveguide structures can also be used for microwave measurements with this housing. Regardless of the particular geometry, the substrate is thermally and electrically anchored to MF. Four adjusting screws are finely tightened for leveling MF with respect to UF and, thereby, for pressing the SMA pins against the film's gold contact pads. This adjustable leveling ensures good electrical contact between the sample and the inner connector pins when using substrates of different thicknesses, which usually are between $0.1$ and $1$~mm. A temperature Cernox and an axial Hall sensor are mounted inside the main frame right underneath the sample~\cite{Lakeshore}. The Hall sensor is used for precise setting small magnetic fields as strong-field solenoids are known for their remanent fields. A nichrome heater wire with a resistance of $40~\Omega$ is enclosed in a separate groove milled around the temperature and the Hall sensors. BL protects the heater and the sensors from undesired mechanical contacts. The bulk copper housing serves as a radiator for providing a uniform temperature distribution during the measurements. In addition to this, the housing electrically shields the sample.

Once the sample is mounted, the top lid closes the housing. The whole construction is screwed onto the cold end of the sample probe. Then the heater and temperature sensing wiring is connected via a Fisher connector to the dc bus. Finally, the six semi-rigid coaxial cables are connected to the mini-SMP slots, electrical contacts of the sample are checked, and the setup is ready for measurements.

\section{Microwave performance}\label{cMicrowave}

To characterize the microwave performance of our setup, we use a thin-film test sample. In general, several planar geometries are suitable for microwave measurements on a thin-film sample. These include a hairpin line, a microstrip line, a coplanar waveguide, and their different modifications. All these geometries can easily be fabricated by photolithographic processes. Aiming primarily at studying the microwave-driven dynamics of Abrikosov vortices in nanopatterned superconductors, we have chosen the microstrip line for the experiment. It is this geometry which allows in the most simple way for adding in-plane leads for measuring voltage drops along and across the sample and thus, for straightforwardly controlling its transition to the normal state. Regardless of the selected geometry, matching of impedances of the bulk and the planar parts of the transmission line is crucial for microwave measurements. Accordingly, to satisfy this condition, the geometrical dimensions of the microstrip have been calculated as follows.

\subsection{Microstrip impedance matching}
\begin{figure}
    \centering
    \includegraphics[width=0.25\textwidth]{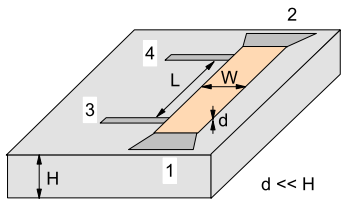}
    \caption{The layout of contacts for measurements of a Nb microstrip. See text for details.
    \label{fMS}}
\end{figure}

The microstrip geometry is sketched in Fig.~\ref{fMS}. The film thickness $d$ is much less than that of the substrate, $d\ll H$, and it is pre-assumed to be between $10$ and $100$~nm. In this case the characteristic impedance of the microstrip line is calculated by the expression~\cite{Poz11boo}
\begin{equation}
\label{eMSI}
    Z_0 = \frac{60}{\sqrt{\varepsilon_e}}\ln\left[\frac{8H}{W} + \frac{W}{4H}\right],\quad \mathrm{for}\quad \frac{W}{H}\leq1,
\end{equation}
where the effective dielectric constant of the microstrip line $\varepsilon_e$ is given approximately by~\cite{Poz11boo}
\begin{equation}
\label{eDCM}
    \varepsilon_e = \frac{\varepsilon_r + 1}{2} + \frac{\varepsilon_r - 1}{2}\frac{1}{\sqrt{1+12H/W}},
\end{equation}
with $W$ being the microstrip width and $\varepsilon_r$ the relative permittivity of the substrate material.

The present experiment assumes the use of Nb films sputtered onto a-cut sapphire substrates with a relative permittivity $\varepsilon_r$ of about $9.3$. Accordingly, to fabricate a 50~$\Omega$-microstrip we had to solve a variational problem in which the ``boundary conditions'' were defined by both, the specifics of the focused ion beam milling (FIB) technique  employed for the fabrication of artificial pinning landscapes and the availability of suitable sapphire substrates. One the one hand, from the viewpoint of the nanoprocessing by FIB, the microstrip geometrical dimensions should be as small as possible. This is due to the resolution limitations of the pattern generator and the pattern drift in the scanning electron microscope (SEM). With our SEM, the latter issue is getting obstructive for accurately writing predefined patterns when the nanoprocessing lasts longer than two hours. On the other hand, the substrates should be thicker than 0.1~mm for the sake of mechanical robustness and handling convenience. Reasonable values we have arrived at are a substrate thickness $H$ of 150~$\mu$m and a microstrip width $W$ of 150~$\mu$m. According to Eqs.~(\ref{eMSI}) and (\ref{eDCM}) the microstrip with these dimensions has an effective dielectric constant $\varepsilon_e = 6.3$ and an impedance $Z_0\approx 50~\Omega$.

\subsection{Sample}
Following the prescription arrived at in the previous subsection, the sample is made from a $50$~nm-thick epitaxial Nb (110) film prepared by dc magnetron sputtering~\cite{Dob12tsf} onto an a-cut ($11\bar{2}0$) sapphire substrate. During the deposition process the substrate was kept at $850^{\circ}$~C, the Ar pressure was $4\times10^{-3}$~mbar, and the growth rate was $\simeq1$~nm/s. Once sputtered, the film was pre-patterned by standard photolithography followed by Ar ion-beam etching in order to define a $50~\Omega$ impedance-matched microstrip with a width $W$ of $150~\mu$m and a length $L$ of $500~\mu$m.

\begin{figure}
    \centering
    \includegraphics[width=0.4\textwidth]{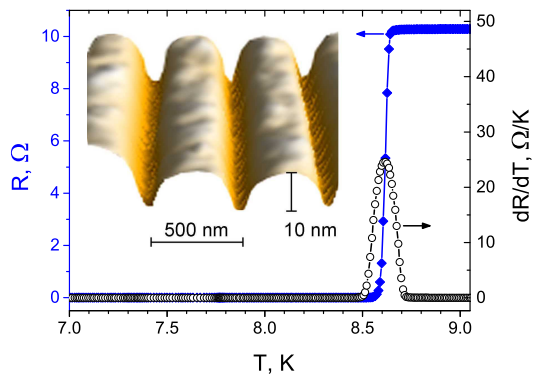}
    \caption[]
    {The superconducting transition of the Nb microstrip patterned with symmetric grooves by focused ion beam milling. Inset: Atomic force microscope image of the microstrip surface.}
   \label{fAFM}
\end{figure}

The microstrip was then nanopatterned by using the focused ion beam milling technique in a high-resolution dual-beam SEM. In the patterning process the beam parameters were $30~$kV/ $50$~pA in normal incidence and the dwell time was $1~\mu$s. The nanopattern is an array of uniaxial straight grooves with a groove-to-groove distance $a$ of $500$~nm, a groove depth of $10$~nm, and a full width at half depth of $80$~nm. The grooves are aligned parallel to the long side of the microstrip, that is parallel to the dc transport current direction. The grooves have a symmetric profile in the cross-section and provide a washboard-like pinning potential landscape for vortices in the mixed state. From our previous experiments we know that the pinning potential induced by milled grooves~\cite{Dob12njp} or, complementary, by deposited Co stripes~\cite{Dob10sst,Dob11pcs} is strong. An atomic force microscope image of the nano-groove array thus fabricated is shown in the inset of Fig.~\ref{fAFM}.

The sample has a room temperature-to-$9$~K resistance ratio of about~$7$ and a residual resistivity $\rho_{\mathrm{9K}}$ of $16~\mu\Omega$cm. The zero-field superconducting critical temperature $T_c$ determined at the transition midpoint is $8.62$~K and the transition width is $\Delta T_c = 0.14$~K, see Fig.~\ref{fAFM}. The sample is characterized by an upper critical field at zero temperature $H_{c2}(0)$ of about $1.3$~T, as deduced from fitting the dependence $H_{c2}(T)$ to the phenomenological law $H_{c2}(T) = H_{c2}(0) (1-t^2)$ \cite{Tin04boo}, where $t = T/T_c$ is the reduced temperature.

\subsection{Setup characterization}

A first characterization of the setup consists in room-temperature measurements of the frequency dependence of the insertion loss of all parts of the microwave transmission line. The purpose is to check whether some section of the line introduces unexpectedly high insertion losses or spurious resonances. The electrical connection is the following: The signal comes from the VNA and the experimentally accessible quantities, from which the absorbed microwave power can be deduced, are the forward $S_{21}$ and the backward $S_{12}$ transmission coefficients. Throughout the text we will use the notation $S_{ij}$ for referring not to the respective coefficient itself, but to its \emph{absolute value}. Thus, $S_{ij}$ is the ratio (expressed in dB) of the microwave power measured at port $j$ to the power transmitted at port $i$. For definiteness, we measure the forward transmission coefficient $S_{21}$ of a standard 50~$\Omega$ feedthrough iteratively included between different elements of the line. Then the 50~$\Omega$ feedthrough is connected as a sample under study and is cooled down to 10~K. The $S_{21}(f)$ curve is a decreasing function of the frequency and does not exhibit sudden variations or modulations (not shown). The measured values of the insertion loss at 1~GHz and 10~GHz are in reasonable agreement with a sum of the nominal losses in the individual transmission line components, as stated by the manufacturers.

\begin{figure}
    \centering
    \includegraphics[width=0.4\textwidth]{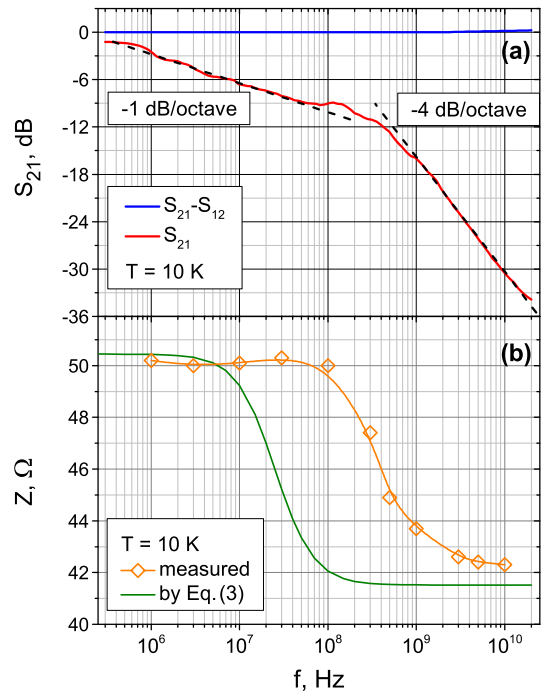}
    \caption{Frequency dependence of (a) the forward transmission coefficient $S_{21}(f)$ and (b) the impedance $Z(f)$ of the sample probe with the mounted Nb microstrip at $10$~K.
    \label{fSvF}}
\end{figure}

The replacement of the feedthrough with the sample-housing box containing the Nb microstrip in the normal state ($T = 10$~K) shifts $S_{21}(f)$ down at frequencies above 500~MHz by 0.3--3~dB below the curve for the 50~$\Omega$ feedthrough, with a qualitatively similar overall shape. The resulting frequency dependence of the forward transmission coefficient at $10$~K is shown in Fig.~\ref{fSvF}(a). The $S_{21}(f)$ trace exhibits attenuation slopes of $-1$~dB/octave and $-4$~dB/octave in the frequency range $1$-$100$~MHz and $1$-$20$~GHz, respectively. The steeper attenuation at elevated frequencies is in part caused by the frequency pass band of the bias-tees and the developing impedance mismatch as deduced from the Smith charts measured for a series of frequencies, refer to Fig.~\ref{fSvF}(b). To interpret the observed reduction of the impedance, we employ a frequency-dependent model for the effective dielectric constant $\varepsilon_e(f)$~\cite[p.\,150]{Poz11boo}
\begin{equation}
\label{eEeF}
    \varepsilon_e(f) = \varepsilon_r - \frac{\varepsilon_r - \varepsilon_e}{1 + G(f)},
\end{equation}
where $\varepsilon_e$ is the effective dielectric constant given by Eq.~(\ref{eDCM}) and $G(f)$, in the form suggested in Ref.~\cite{Bah77mic}, reads $G(f) = g(f/f^\ast)^2$, with $g = 0.6 + 0.009\cdot 50\,\Omega$ and $f^\ast = (50\,\Omega/8\pi)H$, where $f$ is in GHz, and $H$ is in cm. Substituting $\varepsilon_e(f)$ by Eq.~(\ref{eEeF}) into Eq.~(\ref{eMSI}) one arrives at the frequency dependence of the impedance $Z_{f}$ shown in Fig.~\ref{fSvF}(b) which, though quantitatively differs, is capable to qualitatively reproduce the observed reduction of the impedance at elevated frequencies. We attribute the observed difference in the measured and calculated $Z_0(f)$ to the frequency-dependent effects non-captured by the employed model and the much more complicated actual geometry of the sample. In particular, the non-accounted for admittance contributions must stem from (i) the poorly impedance-matched transient SMA pin-to-microstrip areas and (ii) the inclined sections of the microstrip between its parts underneath the SMA pins and the internal $150\,\mu$m$\times 500\,\mu$m area of the microstrip itself.

The difference signal $S_{21}(f)-S_{12}(f)$ plotted in Fig.~\ref{fSvF}(a) allows us to conclude that the $S_{12}(f)$ curve is a replica of the $S_{21}(f)$ trace, within the 0.1~dB uncertainty in the course of repetitive measurements. The shape of the curves in Fig.~\ref{fSvF})(a) remains essentially the same over the broad microwave power range from $-60$~dBm to $+10$~dBm.

\begin{figure}
    \centering
    \includegraphics[width=0.45\textwidth]{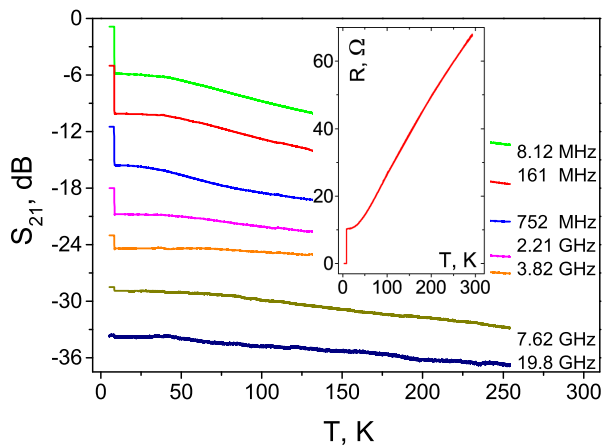}
    \caption{The temperature dependence $S_{21}(T)$ of the transmission line with the mounted Nb microstrip for a series of frequencies, as indicated. Inset: Temperature dependence of the dc electrical resistance of the same Nb microstrip.
    \label{fRvT}}
\end{figure}
The lowest temperature reached with the sample probe is $1.8$~K. When the standard $50\,\Omega$ feedthrough was connected instead of the sample, we observed changes in $S_{21}(T)$ while temperature sweeps from room temperature to $1.8$~K. These changes were small, smooth, and reproducible for all three pairs of coaxial cables. This finding allowed us to save the measured frequency traces in VNA as a reference temperature-sweep-induced background and to present the following temperature dependences with it subtracted.

The $S_{21}(T)$ curves for a series of frequencies measured while cooling down are shown in Fig.~\ref{fRvT}. The complementary temperature dependence of the dc electrical resistance is shown in the inset to Fig.~\ref{fRvT}. One sees that in the lower MHz range the behavior of the $S_{21}(T)$ curves, neglecting the sign, is largely reminiscent of the behavior of the $R(T)$ curve, whereas for the higher GHz frequencies the $S_{12}(T)$ curves become less temperature-dependent and more noisy. This attests to a falling contribution of the superconducting microstrip to the total microwave loss at elevated frequencies. Below $T_c$ the $S_{21}(T)$ exhibit jumps to a less-lossy state, whereas these jumps practically vanish for frequencies in the GHz range. Leaving the observed features in the microwave power for further elaboration, at this point we conclude that the transmission characteristics of the system is free of spurious spikes over the entire accessible temperature range.

\section{Measurement of a microstrip}
\label{cNiobium}
\subsection{Synchronization effects in the current-voltage curves}
\begin{figure}
\centering
    \includegraphics[width=0.48\textwidth]{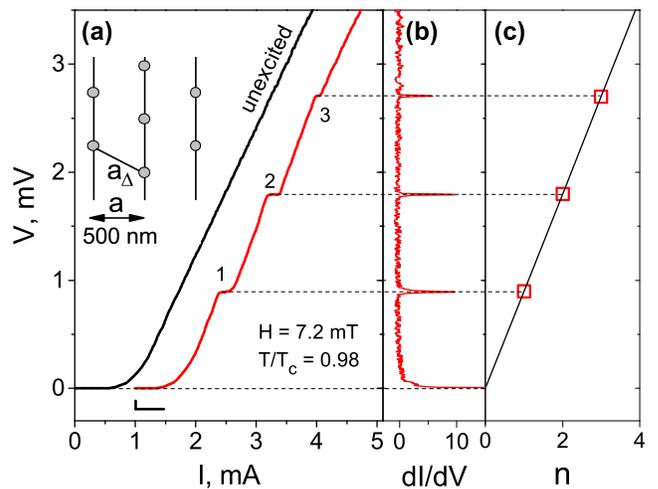}
    \caption[]
    {(a) Manifestation of the Shapiro steps in the flux-flow regime of the CVCs for the microwave stimulus with a frequency of $503$~MHz and a microwave power of $-10$~dBm. The origin of the excited CVC is shifted by $1$~mA to the right along the current axis. Inset: The arrangement of vortices at $7.2$~mT corresponding to the fundamental matching configuration, where $a_\bigtriangleup$ is the vortex lattice parameter and $a$ is the nanopattern period. (b)~The excited CVC in $V-dI/dV$ representation. (c)~Step voltage versus step number for the excited CVC. The solid line is a linear fit to Eq.~(\ref{eCVC}) with $N = 866$.}
   \label{fSteps}
\end{figure}

In this subsection we will show how the dc CVC of the Nb microstrip is modified when a microwave excitation is superimposed onto the dc transport current. The mixed-state section of the unexcited dc CVC of the microstrip measured at the reduced temperature $t = 0.98$ and the fundamental matching field $H = 7.2$~mT is presented in Fig.~\ref{fSteps}. The arrangement of vortices at $7.2$~mT with respect to the underlying pinning nanolandscape for the assumed triangular vortex lattice with lattice parameter $a_\bigtriangleup = (2\Phi_0/B\sqrt{3})^{1/2}$ and the matching condition $a_\bigtriangleup = 2a/\sqrt{3}$ is shown in the inset to Fig.~\ref{fSteps}(a). The CVC is nonlinear, whereby at $I > I_c\approx 0.5$~mA ($I_c$ is determined by the $0.1~\mu$V voltage criterion) one recognizes the linear regime of viscous flux flow with a flux-flow resistivity $\rho_f$ of about $2~\mu\Omega$cm. This is in reasonable agreement with the Bardeen-Stephen expression~\cite{Bar65prv} $\rho_f\simeq 0.9\rho_{\mathrm{9K}} B/B_{c2}(0.98T_c) = 2.1\mu\Omega$cm.

The excited CVC exhibits Shapiro steps occurring at voltages~\cite{Fio71prl}
\begin{equation}
\label{eCVC}
    V = n V_0 \equiv nN\Phi_0 f,
\end{equation}
\begin{figure*}
    \centering
    \includegraphics[width=0.9\textwidth]{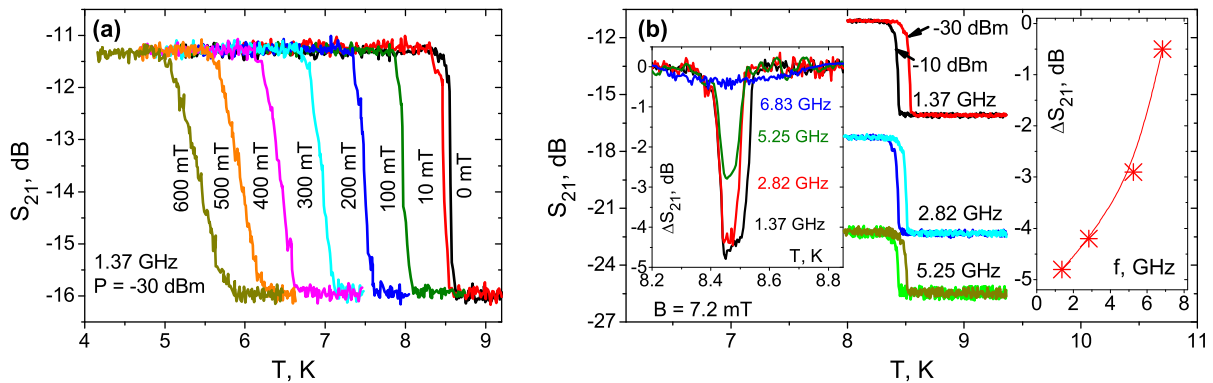}
    \caption{Temperature dependences of the forward transmission coefficient $S_{21}(T)$ measured (a) at $f = 1.37$~GHz and a series of magnetic fields and (b) at $B = 7.2$~mT, two excitation power levels, and a series of  microwave frequencies, as indicated. The left inset in (b) depicts the relative change $\Delta S_{21} = S_{21}$($P =-10$~dBm) $-S_{21}$($P =-30$~dBm), while the right inset displays the excess loss due to the vortex motion as a function of frequency.
    \label{fPower}}
\end{figure*}

where $n$ is an integer (step number), $N$ is the number of vortex rows between the voltage leads, $f$ is the microwave frequency, and $\Phi_0 = 2.07\times10^{-15}$~Vs is the magnetic flux quantum. The steps in the CVCs arise when one or a multiple of the hopping period of Abrikosov vortices coincides with the period of the ac drive. In Fig.~\ref{fSteps}(a) one can distinguish three lowest-order Shapiro steps, whose presence becomes more apparent by plotting $V$ versus $dI/dV$, see Fig.~\ref{fSteps}(b). In Fig.~\ref{fSteps}(c) the step voltage is plotted versus the step number and can nicely be fitted to Eq.~(\ref{eCVC}) with $N=860$-$870$. Given the geometrical dimensions of the microstrip and the fundamental matching field configuration for a triangular vortex lattice in the inset to Fig.~\ref{fSteps}(a) the expected number of vortex rows between the voltage leads is equal to $866$. The fact that the number of vortex rows deduced from fitting the experimental data to Eq.~(\ref{eCVC}) is very close to $866$ suggests that all vortices move coherently. This strongly coherent motion is caused by both, the high periodicity of the nanogroove array and a relatively weak contribution of the background isotropic pinning due to the structural imperfectness as compared to the dominating strong pinning owing to the nanopatterning. This conclusion is in line with our previous observation~\cite{Dob12njp} that the focused ion beam-milled grooves provide a strong pinning potential for vortices forced to move across them. When tuning the field value away from the matching configuration the steps disappear due to lacking coherence in the motion of the vortex lattice. A more detailed analysis of the microwave frequency and power dependences of the depinning current will be reported elsewhere.

\subsection{Microwave power absorption}
\label{sPower}
Another source of important information on the mixed state and the pinning force is the microwave power absorbed by vortices. When the Lorentz force acting on the vortices is alternating, then depending on the microwave frequency two physically different regimes in the vortex motion ensue. The first regime is the localized mode when the vortex is oscillating within one pinning potential trough. In this regime the pinning force dominates and the response is weakly dissipative. This regime ensues at low frequencies. By contrast, at high frequencies the frictional force dominates and the response is strongly dissipative. Thus, the low-dissipative regime is separated from the strongly-dissipative regime by a characteristic crossover frequency which is called the ``depinning frequency'' $f_{dp}$~\cite{Git66prl,Pom08prb} .

The $S_{21}(T)$ curves for the Nb microstrip in the absence of dc current are shown in Fig.~\ref{fPower}(a) for $f = 1.37$~GHz and a series of magnetic fields. At this frequency, the transition to the normal state leads to a reduction of $S_{21}$ by about $5$~dB. With increasing magnetic field the transition broadens and shifts towards lower temperatures, reminiscent of the behavior of the resistance $R(T)$ curves  at the respective fields. In the next measurement, the field is set at the fundamental matching field value $7.2$~mT and the relative changes of $S_{21}(T)$ with the increase of the ac power level by $20$~dB are shown in Fig.~\ref{fPower}(b) for three excitation frequencies. With increasing ac frequency, the microwave loss in the superconducting state increases and the difference between the absorbed power in the superconducting and the normal state is reduced. With reducing the frequency in the microwave range, the relative microwave loss $\Delta S_{21} = S_{21}$($P =-10$~dBm) $-S_{21}$($P =-30$~dBm) saturates at $-5$ dB. Above $7$~GHz the mixed-state losses become nearly as strong as in the normal state. This allows us to estimate the depinning frequency $f_{dp}$ at the $-3$ dB level of $S_{21}$ as $5$~GHz, which is in reasonable agreement with other works~\cite{Jan06prb,Pom10pcs}, taking into account that focused ion beam-milled grooves are very strong linearly-extended pinning sites~\cite{Dob12njp} and the pinning is probed at the groove bottoms, see the upper inset in Fig.~\ref{fSteps}(a), for which elevated depinning frequencies are expected~\cite{Shk12inb,Shk08prb,Shk11prb}. Studying the dependence of the depinning frequency on the dc bias, magnetic field, temperature, as well as sample properties has to remain for future elaboration in forthcoming experiments.

\section{Conclusion}\label{cConclusion}
We have built and successfully tested an insertable cryogenic sample probe for combined microwave and dc electrical transport measurements in a $^4$He cryostat. The setup works from $1.8$~K to room temperature and allows for measurements of the dc and ac voltages and the absorbed microwave power at frequencies from dc to 20~GHz, the high end frequency available with the employed vector network analyzer. Though an attenuation of -30~dB of the microwave signal by the bias tees has been revealed at 10~GHz, the damping of the transmission line itself in conjunction with the power capability of the analyzer allows for microwave measurements at even higher frequencies and is limited by 26~GHz being the high end frequency of the employed SMA connectors. While the inaccessibility of temperatures below 1.8 K appears only as a negligible drawback for measurements on Nb microstrips ($T_c \approx 8.6$~K), for which, in fact, a couple of coaxial cables would suffice, the presence of the six cables is fortunate with regard to further extensions of the setup for more sophisticated experiments.

When measuring the current-voltage curves, we realized that the temperature stability achieved with our bulk sample housing box is within $\Delta T = 1$~mK, which is a factor of three better compared to conventional dc sample probes equipped with the standard sample holder. By the time of writing the sample probe has been used for one year and none of its components have had to be replaced. Further improvements of the experimental setup relate not to the sample probe, but should address the installment of a new two-contour temperature controller, due to a coarse-grain output voltage in the heater circuit of the currently available unit. This will allow for simultaneous tuning the temperature (i) by the heater/sensor in the VTI space and (ii) much more finely, by the heater/sensor mounted in the vicinity of the sample.

\section{Acknowledgements}
OVD thanks (in alphabetical order) F. Aliev, V. Denysenkov, E. Hollmann, K. Il'in, E. Silva, R. W\"orden\-weber, M. W\"unsch, and A. Zaitsev for constructional advices on the sample probe. R. Sachser is thanked for support in automating the data acquisition. Useful discussions with V. A. Shklovskij are acknowledged. This work was supported by the ``Vereinigung von Freunden und F\"orderern der Johann Wolfgang Goethe-Universit\"at'' and the Goethe University funding program ``Nachwuchswissenschaftler im Fokus''. This work was conducted within the framework of the NanoSC-COST Action MP1201.

\end{document}